\documentclass[preprints,accept,moreauthors]{Definitions/mdpi} 
\usepackage{subcaption}

\DeclareMathOperator{\Var}{Var}
\DeclareMathOperator{\sign}{sign}
\DeclareMathOperator{\id}{id}

\firstpage{1}
\pubvolume{1}
\issuenum{1}
\articlenumber{0}
\pubyear{2021}
\copyrightyear{2021}
\externaleditor{Academic Editor: Firstname Lastname} % For journal Automation, please change Academic Editor to "Communicated by"
\datereceived{}
\dateaccepted{}
\datepublished{}
\hreflink{https://doi.org/} % If needed use \linebreak
\Title{Wealth rheology}

\Author{Zdzislaw Burda $^{1,\dagger}$\orcidA{}, Malgorzata J. Krawczyk $^{1,\dagger}$\orcidB{}, Krzysztof Malarz $^{1,*,\dagger}$\orcidC{}, and Malgorzata Snarska $^{2,\dagger}$\orcidD{}}

\AuthorNames{Zdzislaw Burda, Malgorzata J. Krawczyk, Krzysztof Malarz and Malgorzata Snarska}

\address{$^{1}$ \quad AGH University of Science and Technology, Faculty of Physics and Applied Computer Science, Mickiewicza 30, PL-30059 Krak\'ow, Poland;\\
$^{2}$ \quad Cracow University of Economics, Department of Financial Markets, Rakowicka 27, PL-31510 Krak\'ow, Poland
}

\corres{Correspondence: malarz@agh.edu.pl}

\firstnote{These authors contributed equally to this work.} 
\abstract{We study wealth rank correlations in a simple model of macroeconomy.
To quantify rank correlations between wealth rankings at different times,
we use Kendall's $\tau$ and Spearman's $\rho$, Goodman--Kruskal's $\gamma$, 
and the lists' overlap ratio. We show that the dynamics of wealth flow 
and the speed of reshuffling in the ranking list depend on parameters of the
model controlling the wealth exchange rate and the wealth growth volatility. 
As an example of the rheology of wealth in real data, we analyze the lists of the 
richest people in Poland, Germany, the USA and the world.}


\begin{document}
%%\linenumbers

%% ####################################################################################
\section{Introduction}
%% ####################################################################################

The problem of wealth inequality is the subject of intense research
in economics \cite{Davis_2000,Benhabib_2016,De_Nardi_2015,Berman_2016a,Berman_2016b}, sociology and econophysics \cite{chakrabarti_chakraborti_chakravarty_chatterjee_2013}, 
but it also arouses great interest outside of science \cite{Oxfam_2017,WIR_2018,Piketty_2014,Piketty_2015,Wprost,Spiegel,Forbes}. 
Parallel to empirical studies, theoretical research is carried out to explain the main features of wealth statistics and wealth dynamics observed in macroeconomic data, like the presence of Pareto tails in wealth distribution or increasing wealth inequality in the world.

In theoretical models, wealth dynamics is often described by stochastic equations representing the evolution of wealth of a typical individual 
in a given economic environment. In physical terminology, this can be called the one-body approach. Even today this approach is used in mainstream economics, for example, in studies of wealth inequality \cite{Piketty_2012}. An alternative is to use population dynamics, 
which in this context is often referred to as agent-based modelling. 
It was introduced in \cite{Angle_1986} and popularised in \cite{Samanidou_2007}. 
In agent-based modelling, the economy is perceived as a complex system consisting of many entities interacting with each other under given macroeconomic conditions. Unlike the one-body approach, this can be called the multi-body approach.
The main idea of agent-based modelling is to statistically look at the problem of wealth distribution from the perspective of the entire system.
This allows studying collective effects, like correlations between agents, or emergent phenomena, such as the formation of wealth classes, or self-organisation of the economy, or instability of the system. 
This perspective is in many aspects similar to that used in statistical physics, aiming at deriving macroscopic physical laws from microscopic rules by applying laws of large numbers. This is probably why the problem of wealth distribution has been intensively studied in econophysical literature
\cite{chakrabarti_chakraborti_chakravarty_chatterjee_2013}.
Many ideas behind agent-based modelling have been derived from concepts like kinetic
theory \cite{Dragulescu_2000,Yakovenko_2009}, scattering \cite{Slanina_2004}, 
rate equations \cite{Ispolatov_1998}, 
random matrix theory \cite{Burda_2003,Bouchaud_2018}, 
Brownian motion \cite{Van_Kampen_2007} which were developed
in statistical physics. Using this type of ideas, one was able to
model wealth or income distributions \cite{Bouchaud_2000}, 
dynamics of wealth inequality \cite{Adamou_2016,Burda_2019}, wealth
concentration \cite{Snarska_2020}, structure emergence \cite{Wasko_2006,Karpinska2004b}, 
economic instability and corruption mechanisms \cite{Burda_2002,Ball_2002,Kulakowski2016}, 
systemic risk in economic networks \cite{Thurner_2020},  
emergence of heavy tails in wealth and income distributions 
\cite{Levy_1996,Bouchaud_2000}, 
and herding behaviour \cite{Cont_2000}, or to analyse statistical 
behaviour or rational agents \cite{VENKATASUBRAMANIAN2015120}. 

Since the time of Keynes, it has been widely believed that a closed economy eventually reaches a stationary state, 
also known as a steady state or saturation.
What is usually meant as a stationary state economy is a system where macroeconomic quantities have stationary distributions.
A typical example is the wealth distribution that does not change over time after reaching a steady state.
This does not mean that each person's wealth is constant in time, but that the system 
as a whole has a stationary distribution in a statistical sense. In fact, the wealth of individuals may change all the time even in the stationary state. Wealth flows from one individual to another: some people get richer, some poorer, {\it panta rhei}. In this article, 
we will take a closer look at the flow of wealth. We call this class of phenomena {\em wealth rheology}. To be specific, in the paper we study the dynamics of wealth rank correlations 
using the Bouchaud--M\'ezard model of macroeconomy \cite{Bouchaud_2000}.
The model is implemented as a stochastic process based on Gibrat's law of proportionate growth \cite{Gibrat_1931} which is combined with dynamics representing agents' interactions. The model generates Pareto's tail in the wealth distribution \cite{Pareto_1897}. The stochastic process belongs to the class of Kesten processes \cite{Kesten_1973}, 
which is a class of multiplicative contracting stochastic processes. 
It is a generic feature of Kesten processes that they lead to a stationary state with a power-law tail.

%% ####################################################################################
\section{The model}
%% ####################################################################################

In this section we briefly recall the  Bouchaud--M\'ezard model \cite{Bouchaud_2000}.
The model describes evolution of wealth of $N$ interacting agents in a closed macroeconomic system. 
The evolution is given by $N$ stochastic differential equations  
for wealth $W_a(t)$ of agents $a=1,\ldots,N$  at time $t$.
In the continuous time formalism the equations read
\begin{equation}
\frac{dW_a(t)}{dt} = 
\left[\left(\mu_* + \frac{1}{2} \sigma_*^2\right) + \sigma_* \frac{dB_a(t)}{dt}\right] W_a(t) +
\sum_{b=1}^N \left( J_{ab} W_{b}(t) - J_{ba} W_a(t) \right),
\label{BM}
\end{equation}
where $B_a(t)$, $a=1,\ldots,N$ are independent
Wiener processes (continuous Brownian motions).
The above equations are written in the It\^{o} formalism\footnote{In the Stratonovich approach the term $\mu_*+\frac{1}{2}\sigma_*^2$ would be replaced with $\mu_*$.}. 
In this paper we shall use 
the discrete time formalism. Equation (\ref{BM}) can be discretised by introducing an elementary time interval $\Delta t$ relating physical time $t$ to discrete time 
$k=0,1,2,\ldots$ as follows $t= k \Delta t$.  
In the leading order, up to $\mathcal{O}(\Delta t)$-terms, 
the discretisation of  Equation (\ref{BM}) gives
\begin{equation}
W_{a,k} = \exp\left(r_{a,k}\right) W_{a,k-1}  + 
\sum_{b=1}^N \left( j_{ab} W_{b,k-1} - j_{ba} W_{a,k-1} \right),
\label{BMD}
\end{equation}
where $W_{a,k} \equiv W_a(t)$, $j_{ab} \equiv J_{ab} \Delta t$ and 
$r_{a,k}$ are independent identically distributed random variables with the normal distribution $r_{a,k} \sim \mathcal{N} (\mu, \sigma^2)$ with 
$\mu = \mu_* \Delta t$ and $\sigma = \sigma_* \sqrt{\Delta t}$. 
In the limit $\Delta t \rightarrow 0$, Equation \eqref{BM} is restored. 
The first term on the right hand side of
Equation \eqref{BMD} corresponds to random multiplicative fluctuations of wealth.
In economic literature it is referred to as the law of proportionate effect \cite{Gibrat_1931} stating that growth rates $r_{a,k}$ are independent of wealth. 
In the simplest version of the model it is assumed that the growth rates have 
the same mean ${\rm E}(r_{a,k})=\mu$ and  the same variance  
$\Var(r_{a,k})=\sigma^2$ for all agents throughout evolution of the system. 
The first term on the right hand side of Equation (\ref{BMD}) reflects spontaneous changes in wealth due to changing market conditions and expectations. The second one describes the flow of wealth between 
individuals, resulting from interactions and trading. Coefficient $j_{ab}$ is 
the fraction of wealth of agent $a$ which is transferred to agent $b$ within a single 
time interval $\Delta t$. Equations \eqref{BMD} are invariant under rescaling of wealth 
by a common factor  $W_{a,k} \rightarrow W'_{a,k}=  \lambda W_{a,k}$ for all $a=1,\ldots,N$.
In particular, this means that the equations do not change when the monetary units change. Therefore, it is convenient to express wealth in units of the mean wealth $\overline{W}_k = \frac{1}{N} \sum_{a=1}^N W_{a,k} $. We denote the corresponding quantities by small letters
\begin{equation}
w_{a,k} = \frac{W_{a,k}}{\overline{W}_k}.
\label{nw}
\end{equation}
The normalised wealth values \eqref{nw} 
are insensitive to the parameter $\mu$ controlling 
the mean growth rate, because it cancels in
the numerator and the denominator of Equation (\ref{nw}). 
The change $\mu \rightarrow \mu'=\mu + \Delta \mu$
can be interpreted as a change of the inflation rate by $\Delta \mu$.

The model can be solved in the mean-field approximation assuming that
all agents interact with each other with the same intensity
$J_{ab} = \frac{J}{N}$ for all pairs $a\ne b$.  For $J>0$, in the limit $N\to\infty$ and $\Delta t \to\infty$, the normalised wealth \eqref{nw} can be shown to approach a stationary state with the distribution given by the 
inverse gamma distribution with the following probability density function \cite{Bouchaud_2000}:
\begin{equation}
p_{\text{eq}}(w) = \frac{(\alpha-1)^\alpha}{\Gamma(\alpha)}
\frac{\exp\left( -\frac{\alpha-1}{w}\right)}{w^{1+\alpha}},
\label{peq}
\end{equation}
where the parameter $\alpha$ is
\begin{equation}
\alpha = 1 + \frac{2J}{\sigma_*^2} = 1 + \frac{2j}{\sigma^2}.  
\label{alpha}
\end{equation}
One can easily check that the mean of this distribution is equal to one, that is
${\rm E}(q) = \int_0^\infty w p_{\text{eq}}(w) dw=1$ in accordance with the normalisation \eqref{nw}. The distribution has a Pareto tail 
$p_{\text{eq}}(w) \sim w^{-1-\alpha}$ for $w \gg 1$ with the exponent $\alpha$, given by Equation \eqref{alpha}. 
The index $\alpha$ depends on the ratio of the flow 
intensity parameter $j$ and the volatility of growth rates $\sigma^2$,
so the stationary distribution does not change when $\sigma^2$ and $j$ are simultaneously re-scaled by the same factor $\sigma^2 \rightarrow \sigma'^2 =\lambda \sigma^2$ and $j\rightarrow j'=\lambda j$.
The parameters $\sigma$ and $j$ can be interpreted as economic activity parameters. 
The flow parameter $j$ reflects the intensity of trade 
and wealth exchange. The parameter $\sigma$ is the growth
rate volatility and it reflects the degree of economic freedom: the larger $\sigma$ the more
liberal economy. Large values of $\sigma$ mean that the state does not intervene 
and does not help if economic entities need support. On the contrary
it supports and encourages a free market, new ideas, bold businesses and the foundation of start-ups, etc. In effect, some companies may quickly grow, while some large established companies can shrink or go bankrupt quickly. For a large $\sigma$, large changes in the wealth of individuals are expected. In such circumstances, the economic landscape is changing rapidly.
On the other hand, small values of $\sigma$ mean that the economy is very conservative, 
that is, the system discourages risky investments and the state intervenes when established companies need help, the system supports economic \emph{status quo} and the economy is more predictable in the short term.

The aim of this paper is to compare the wealth dynamics in steady state for systems having the same stationary distribution \eqref{peq}, but different economic activity parameters $\sigma$ and $j$. To get an insight into wealth flow dynamics in 
the steady state, we study temporal evolution of wealth rank correlations 
and quantify them by measuring Kendall's $\tau$ \cite{Kendall} and Spearman's $\rho$  \cite{Spearman} for wealth distributions separated by $k$ steps of evolution \eqref{BMD}. 
Standard definitions of rank correlations are recalled in Appendix \ref{AppA}.

%% ####################################################################################
\section{Monte Carlo simulations}
%% ####################################################################################

We perform Monte Carlo simulations to generate evolution of the system.
In practice we find it more convenient to use a slightly modified version of 
the evolution equations \eqref{BMD}, where a single step of evolution is split
into two 
\begin{equation}
\begin{split}
W_{a,k-\frac12} & = \exp\left(r_{a,k}\right) W_{a,k-1},\\
W_{a,k} & =  W_{a,k-\frac12} + \sum_{b=1}^N \left( j_{ab} W_{b,k-\frac12} - j_{ba} W_{a,k-\frac12} \right)
\end{split}
\label{BMD2}
\end{equation}
with some intermediate values $W_{a,k-\frac12}$. 
The first equation in \eqref{BMD2} 
corresponds to Gibrat's rule of proportionate growth,
while  the second one to wealth's flow which preserves the total
wealth in the system $\sum_{a=1}^N W_{a,k}=\sum_{a=1}^N W_{a,k-\frac12}$.
One can easily show that the two representations of the evolution \eqref{BMD}, \eqref{BMD2} 
are identical up to $\mathcal{O}(\Delta t)$-order so they
have the same continuous time limit
(\ref{BM}) for $\Delta t \rightarrow 0$. 
Here, for the sake of simplicity, we focus on the mean field system where each agent interacts with all the others with the same intensity $j_{ab} = j/N$. In this case, Equations \eqref{BMD2} simplify to
\begin{equation}
\begin{split}
W_{a,k-\frac12} & = \exp\left(r_{a,k}\right) W_{a,k-1},\\
W_{a,k} & = (1 - j) W_{a,k-\frac12} +  \frac{j}{N} \sum_{a=1}^N W_{a,k-\frac12}.
\end{split}
\label{mfe}
\end{equation}
The flow rate $j$ can be interpreted as the average fraction of wealth that can flow from an agent to others over a period of time $\Delta t$.

We simulated systems up to $N=10^6$ agents, but the results presented in this paper are for $N=10^4$.
We used two types of initial configurations:
\begin{itemize}
\item a complete equality configuration where $W_{a,0}=1$ for all $a=1,\ldots,N$;
\item an equilibrium configuration, where $W_{a,0}$ are drawn independently of each other from the inverse gamma distribution \eqref{peq}.
\end{itemize}
We call them `cold' and `hot' starts, respectively.

%% ####################################################################################
\section{Results}
%% ####################################################################################

\begin{figure}[htb]
\centering
%% ------------------------------------------------------------------------------------
\begin{subfigure}[b]{0.49\columnwidth}
\caption{\label{fig:Gini_a}$G(\alpha)$}
\vspace{2mm}
\centering
\includegraphics[width=0.99\textwidth]{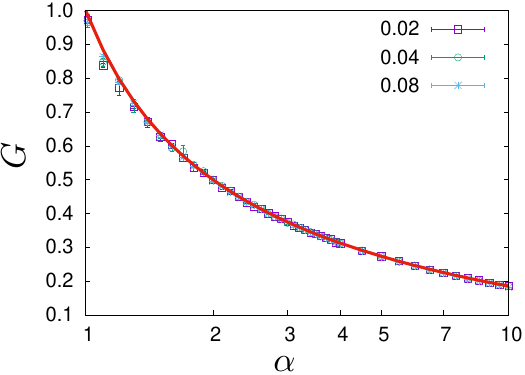}
\end{subfigure}
%% ------------------------------------------------------------------------------------
\begin{subfigure}[b]{0.49\columnwidth}
\caption{\label{fig:Gini_b}$\tau_{\text{ac}}(\sigma)$ and $\tau_{\text{exp}}(\sigma)$}
\vspace{2mm}
\centering
\includegraphics[width=0.99\textwidth]{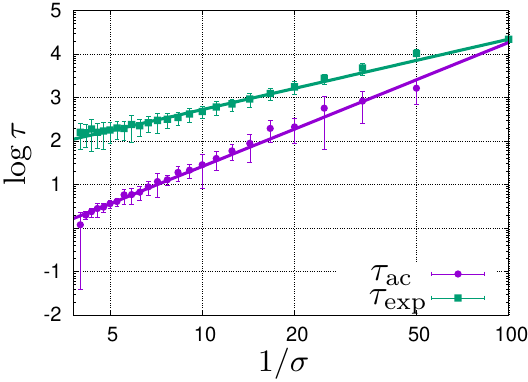}
\end{subfigure}\\
%% ------------------------------------------------------------------------------------
\begin{subfigure}[b]{0.49\columnwidth}
\caption{\label{fig:Gini_c}$G$, the `cold' start}
\vspace{2mm}
%%\centering
\includegraphics[width=0.99\columnwidth]{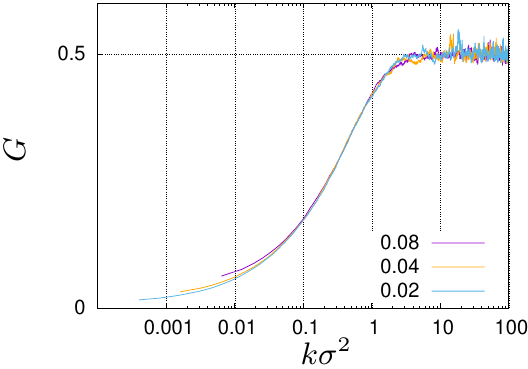}
\end{subfigure}
%% ------------------------------------------------------------------------------------
\begin{subfigure}[b]{0.49\columnwidth}
\caption{\label{fig:Gini_d}$G$, the `hot' start}
\vspace{2mm}
%%\centering
\includegraphics[width=0.99\columnwidth]{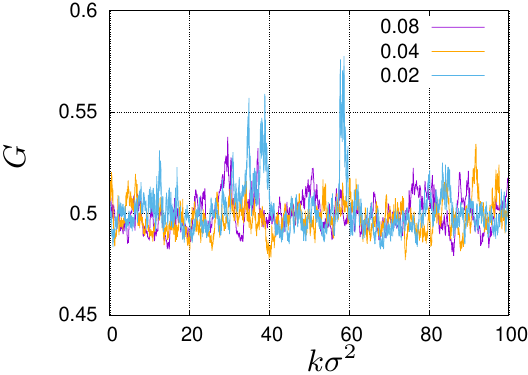}
\end{subfigure}
%% ------------------------------------------------------------------------------------
\caption{\label{fig:Gini}
(a) Gini coefficient $G$ \eqref{Galpha} plotted as a function of $\alpha$ (solid line) and computed numerically from samples generated in Monte Carlo simulations for $N=10^4$ (symbols).
Different symbols correspond to $\sigma=0.02$, $0.04$ and $0.08$.
(b) The auto-correlation time (1) $\tau_{\text{ac}}$ and the exponential time (2) $\tau_{\text{exp}}$ for the Gini coefficient $G$ measured for consecutive configurations in the stationary state for $\alpha=2$ for different $\sigma$.
When $\sigma$ decreases $\tau_{\text{ac}}$ grows as $\sigma^{-x}$ with $x=2.849(50)$, and $\tau_{\text{exp}}$ grows as $\sigma^{-y}$ with $y=1.617(15)$.
(c) Evolution of the Gini coefficient $G$ from the `cold' start, $G=0.0$, towards the stationary state's value $G=0.5$ for $\alpha=2.0$.
The plots correspond to $\sigma=0.02$, $0.04$ and $0.08$. Please note logarithmic scale on the time axis. 
(d) Evolution of the Gini coefficient $G$ from the `hot' start. The values fluctuate about the stationary state value $G=0.5$ for $\alpha=2.0$.
The plots correspond to $\sigma=0.02$, $0.04$ and $0.08$.}
%% ------------------------------------------------------------------------------------
\end{figure}

In Figure \ref{fig:Gini_a} we compare a theoretical prediction for the Gini coefficient 
with the values obtained in Monte Carlo simulations of the system with $\sigma=0.02,0.04,0.08,$ and $N=10^4$. 
The theoretical prediction for the distribution \eqref{peq} reads \cite{Burda_2019}
\begin{equation}
G(\alpha) =\frac{\Gamma(2\alpha-1)}{\Gamma(\alpha)} 
\left\{ \frac{_2F_1(\alpha-1,2\alpha-1;\alpha;-1)}{\Gamma(\alpha)} + 
\frac{(1-\alpha) \, _2F_1(\alpha,2\alpha-1;\alpha+1;-1)}{\Gamma(\alpha+1)}\right\},
\label{Galpha}
\end{equation}
where $_2F_1(a,b;c;z)$ is the hypergeometric function \cite{Arfken}. One can see in Figure \ref{fig:Gini_a} 
that the experimental and theoretical values are consistent. This means that the evolution equations \eqref{mfe} 
bring the system to the predicted steady state. In fact, there are some slight deviations from the theoretical prediction
which can be attributed to the fact that the theoretical
results are derived in the continuous time formalism, while the Monte Carlo simulations are done for discrete time. 
The deviations grow with the volatility $\sigma$.

%% ====================================================================================
%% ====================================================================================

The main conclusion from the comparison shown in Figure \ref{fig:Gini_a} is that the stationary state in the first-order approximation depends on the combination $j/\sigma^2$ and not on $\sigma$ itself, 
exactly as predicted by the theoretical formula \eqref{Galpha}. Let us now address the question how
the dynamic properties of evolution depend on $\sigma$. First, we will study the rate of relaxation towards the  steady state by measuring the exponential auto-correlation time $\tau_{\text{exp}}$ \cite{Madras_1988} for the Gini coefficient. Roughly speaking
the exponential time corresponds to the time needed for the system to reach the stationary state.
To measure $\tau_{\text{exp}}$ we initiate the system from a uniform wealth distribution (cold start),
for which the Gini coefficient is $G=0$, and wait till the Gini coefficient of
the current configuration exceeds the steady state value for the first time. In Figure \ref{fig:Gini_b} we
show the mean exponential auto-correlation time $\tau_{\text{exp}}$ as a function of $\sigma$ for $\alpha=2$ 
and $N=10^4$ averaged over a sample of $100$ values 
obtained from independent simulations.
We see that $\tau_{\text{exp}}$ grows as $\sigma$ decreases. This
effect is clearly seen in Figure \ref{fig:Gini_c} which shows
examples of the evolution of the Gini coefficient for $N=10^4$, $\alpha=2$ from cold starts for $\sigma=0.02$, $\sigma=0.04$
and $\sigma=0.08$. In all three cases 
the Gini coefficient evolves from the initial value $G=0$ towards the stationary state 
value \eqref{alpha} which is equal to $G=0.5$ for $\alpha=2$. The values of $G$ during the evolution are plotted against $k\sigma^2$ in Figure \ref{fig:Gini_c}. The variable $k\sigma^2$ on the horizontal axis is proportional to physical time $t = k\Delta t$, for given $\sigma_*$ in Equation (\ref{BM}), because $\sigma^2 = \sigma_*^2 \Delta t$. The curves in Figure \ref{fig:Gini_c} correspond to different time intervals $\Delta t$ used in the discretization of Equation (\ref{BM}). The small variations between the curves, seen in Figure \ref{fig:Gini_c} for small  $k \sigma^2$, can be attributed to the higher-order corrections in $\Delta t$ that occur in the discretization of Equation (\ref{BM}), skipped in Equation (\ref{BMD}). Generally, we can expect that in the presented range of $\sigma$ the evolution is universally described by the parameter $k \sigma^2$, that we shall use from here on.

Once the stationary state is reached, the value of the Gini coefficient fluctuates about 
the steady state value. This is illustrated in Figure \ref{fig:Gini_d} where we show evolution of the Gini
coefficient in systems initiated from hot starts, that correspond to the stationary
wealth distribution given by Equation (\ref{peq}). 
The Gini coefficient fluctuates about the stationary value $G=0.5$
but the way it fluctuates about this value slightly depends on the volatility $\sigma$. The reason for this is related to 
the presence of a heavy tail in the limiting distribution (\ref{peq}), which for $\alpha=2$, leads to an infinite variance. 
In effect, the curves representing the evolution in Figure \ref{fig:Gini_d} exhibit strong fluctuations which depart from mean value. 
For the simulations shown in the figure, the length of evolution measured in the discrete time steps, $k$, is sixteen and four times longer for 
$\sigma=0.02$ and $\sigma=0.04$, respectively, than for $\sigma=0.08$. Therefore, we observe more fluctuations and larger deviations 
for $\sigma=0.02$ and $\sigma=0.04$ than for $\sigma=0.08$. To quantify the degree of correlation between the values of the Gini coefficient
at different times, we measure the integrated auto-correlation time $\tau_{\text{ac}}$ 
\cite{Madras_1988} which gives a typical timescale for the length of temporal fluctuations. Coming back to
Figure \ref{fig:Gini_b} we show there the dependence of the integrated auto-correlation time $\tau_{\text{ac}}$ on 
$\sigma$ for the system with $\alpha=2$ and $N=10^4$.
It can be compared in the figure to the relaxation time $\tau_{\text{exp}}$ that we previously discussed.
As you can see, both $\tau_{\text{ac}}$ and $\tau_{\text{exp}}$ grow when $\sigma$ decreases. The straight lines in Figure \ref{fig:Gini_b} are 
to guide the eye. They are determined as the best power-law fits:
$\tau_{\text{ac}} \propto \sigma^{-x}$, where $x=2.849(50)$ and
$\tau_{\text{exp}} \propto \sigma^{-y}$, where $y=1.617(15)$.

As mentioned in the introduction, we are mainly interested in the dynamic aspects of the evolution of wealth distribution, such as wealth flow and wealth ranking reshuffling.
We are now going to analyse the wealth rank correlations for
different systems having the same stationary state, 
by studying systems with different $\sigma$'s and $j$'s but identical $\alpha$'s \eqref{alpha}.

\begin{figure}[ht]
\centering
%% ------------------------------------------------------------------------------------
\begin{subfigure}[b]{0.49\columnwidth}
\caption{\label{fig:rank_correlations_a}}
\vspace{2mm}
\includegraphics[width=0.99\textwidth]{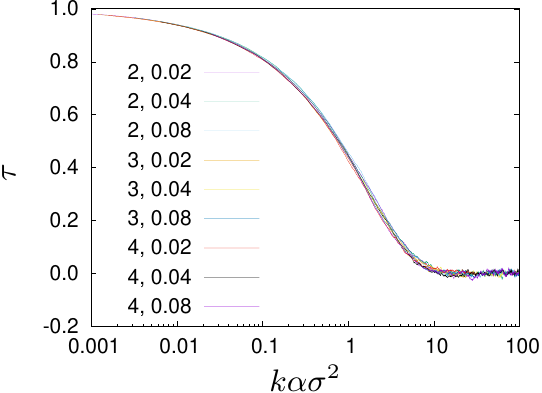}
\end{subfigure}
%% ------------------------------------------------------------------------------------
\begin{subfigure}[b]{0.49\columnwidth}
\caption{\label{fig:rank_correlations_b}}
\vspace{2mm}
\includegraphics[width=0.99\textwidth]{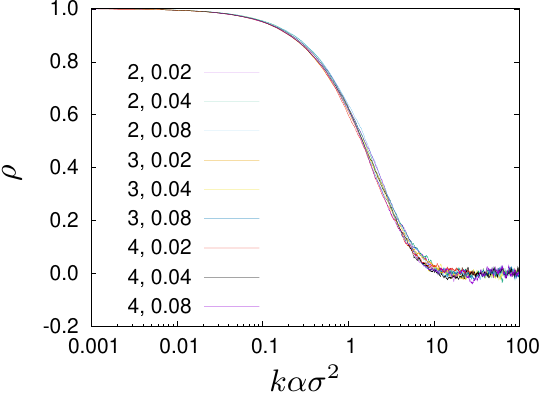}
\end{subfigure}
%% ------------------------------------------------------------------------------------
\caption{\label{fig:rank_correlations} Rank correlations coefficients   
(a) $\tau$ and  
(b) $\rho$ for steady state configurations separated by $k$ time steps in the simulated systems for 
$N=10^4$ and for various values of $\alpha=2$, $3$, $4$ and for various values of $\sigma=0.02$, $0.04$ and $0.08$.
Please note logarithmic scale on the time axis.
The first parameter in the legend is the value of $\alpha$ and the second is the value of $\sigma$.} 
\end{figure}

Wealth rank is a position on the ranking list ordered from the richest to the poorest individuals. The wealth ranking is continuously reshuffling, even in the stationary state. The rate of reshuffling depends on $\sigma$ and $j$: the larger $\sigma$ and $j$ the faster is the reshuffling. Changes in wealth ranking are faster when $\sigma$ is larger.
In Figure \ref{fig:rank_correlations} we show rank correlations between wealth ranking lists obtained in the Monte Carlo
simulations of the Bouchaud--M\'ezard model \cite{Bouchaud_2000}, for configurations separated by 
$k$ steps of evolution \eqref{mfe}, in the stationary 
state of the system with $N=10^4$, and $\alpha=2.0$, $3.0$ and $4.0$, and $\sigma=0.02$, $0.04$ and $0.08$. The degree of rank correlations is quantified by Kendall's $\tau$ \cite{Kendall} and Spearman's $\rho$ \cite{Spearman}
$\rho$ (see Appendix \ref{AppA}). In Figure \ref{fig:rank_correlations} we plot $\tau$ and $\rho$ against a rescaled variable $k\alpha\sigma^2$. 
We can see that the curves lie on top of each other, reflecting some universality of the wealth rheology in the model (\ref{BM}). 

%% ====================================================================================
%% ====================================================================================

\begin{figure}[ht]
\centering
\includegraphics[width=0.49\columnwidth]{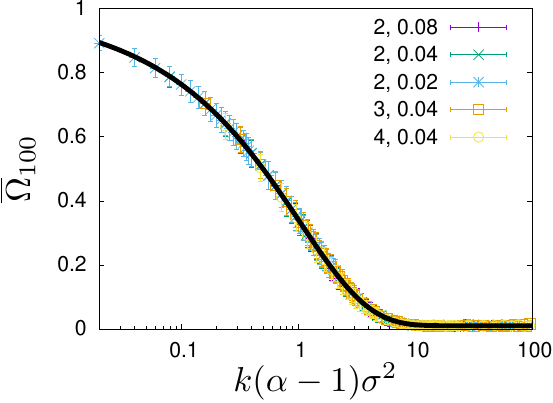}
\caption{\label{fig:top100-decay}
Dependence of the overlap of top-100 lists at times $k_1$ and $k_2$ on the separation time $k=k_2-k_1$. 
The overlap is measured as the percentage of people that are on both the lists. 
The data points are obtained by averaging over pairs of $k_1$, $k_2$ such that $k_2-k_1=k$. 
They are plotted against the universal argument $x=k\sigma^2(\alpha-1)$. The data is obtained by simulations of the model for $N=10^4$, and for different combinations
of $\alpha=2$, $3$, $4$ and $\sigma=0.02$, $0.04$, $0.08$.
The first parameter in the legend is the value of $\alpha$ and the second is the value of $\sigma$.
The data is fitted with the formula \eqref{eq:omega} 
%% $\overline{\Omega}_{100}(k\sigma^2)=0.99\exp(-A\sqrt{k\sigma^2}-Bk\sigma^2)+0.01$
with $A=0.7570(26)$ and $B=0.3341(30)$. The fit is shown with a solid line.}
%% Final set of parameters            Asymptotic Standard Error
%% =======================            ==========================
%% aom             = 0.756965         +/- 0.00262      (0.3461%)
%% bom             = 0.334122         +/- 0.003016     (0.9025%)`
\end{figure}

Another quantity which captures rank correlations is the
overlap ratio which is defined as the percentage of people 
which are among $n$ richest people at times $k_1$ 
and $k_2$.
For example, you may be interested in how many people from the top-100 richest list in some year, are in the top-100 richest list one or two years later.
If $T_n(k_1)$ denotes the set of people being in the top-$n$ richest list in the ranking at time $k_1$ and $T_n(k_2)$ at time $k_2$, the overlap ratio is:
\begin{equation}
\Omega_n(k_1,k_2) = \frac{\# \left(T_n(k_1) \cap T_n(k_2)\right)}{n}
\end{equation}
where the symbol $\cap$ denotes sets' intersection, and
$\#$ --- set's cardinality. If the dynamics describing the
wealth evolution is Markovian, then the overlap ratio 
$\Omega_n(k_1,k_2)$ depends on the time difference $k=k_2-k_1$. 
In such a case, the overlap ratio can be estimated 
numerically as follows 
\begin{equation}
\overline{\Omega}_n(k) = \frac{1}{K-k} \sum_{j=1}^{K-k} \Omega_n(j,j+k),
\label{overlap_ratio}
\end{equation}
where $k$ is the discrete time, which 
refers to consecutive configurations (rankings),
and $K$ is the number of configurations in the sample.
In Figure \ref{fig:top100-decay} we show the expected overlap $\overline{\Omega}_{100}$ for 
the top-100 richest list, estimated from the steady state configurations in Monte Carlo simulations for 
three values of $\alpha$ \eqref{peq} and for three values of $\sigma$, for $N=10^4$. We see that all curves collapse
to a single universal curve. By listing explicitly all arguments of the overlap ratio $\Omega_n(k,\sigma,\alpha,N)$,
we see that the overlap ratio becomes a universal function of the argument $x=k\sigma^2(\alpha-1)$:
\begin{equation}
    \Omega_n(k,\sigma,\alpha,N) = \omega_n\left(x=k\sigma^2(\alpha-1),N\right).
    \label{eq:scaling}
\end{equation}
The function $\omega_n(x,N)$ interpolates monotonically between $\omega_n(x,N)= 1$ for $x\rightarrow 0$ and $\omega_n(x,N) = \frac{n}{N}$ for
$x\rightarrow \infty$.
A simple phenomenological formula
\begin{equation}
    \omega_n(x,N) = \left(1-\frac{n}{N}\right) e^{-A \sqrt{x} - B x} + \frac{n}{N}
\label{eq:omega}
\end{equation}
gives a very good fit to the data (see Figure \ref{fig:top100-decay}). Notice that the scaling variable $k\sigma^2(\alpha-1)$ is not the same as the scaling variable $k\alpha\sigma^2$ for $\rho$ and $\tau$.

%% ===================================================================================
%% ====================================================================================

\begin{figure}[ht]
%% ------------------------------------------------------------------------------------
\begin{subfigure}[b]{0.49\columnwidth}
\caption{\label{fig:real_data_www_poland}Poland, data from \cite{Wprost}}
\vspace{2mm}
\includegraphics[width=0.99\columnwidth]{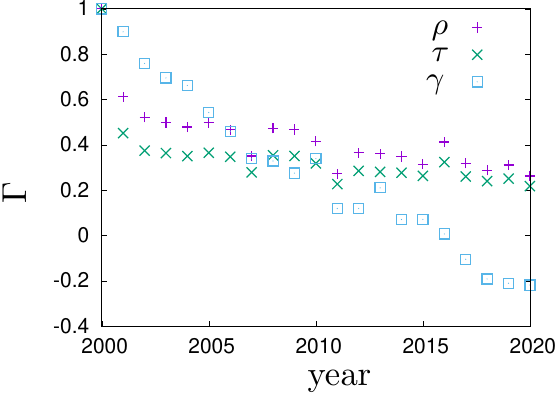}
\end{subfigure}
%% ------------------------------------------------------------------------------------
\begin{subfigure}[b]{0.49\columnwidth}
\caption{\label{fig:real_data_germany}Germany, data from \cite{Spiegel}}
\vspace{2mm}
\includegraphics[width=0.99\columnwidth]{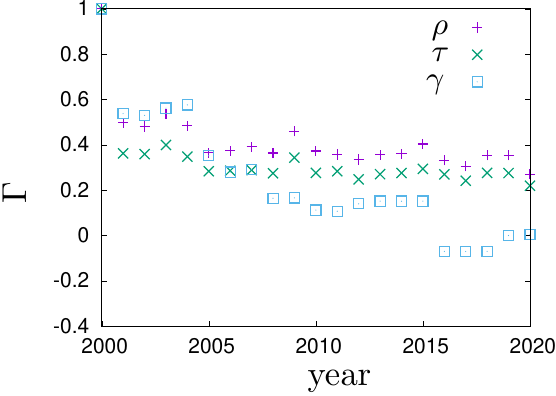}
\end{subfigure}\\
%% ------------------------------------------------------------------------------------
\begin{subfigure}[b]{0.49\columnwidth}
\caption{\label{fig:real_data_usa}the USA, data from \cite{Forbes}}
\vspace{2mm}
\includegraphics[width=0.99\columnwidth]{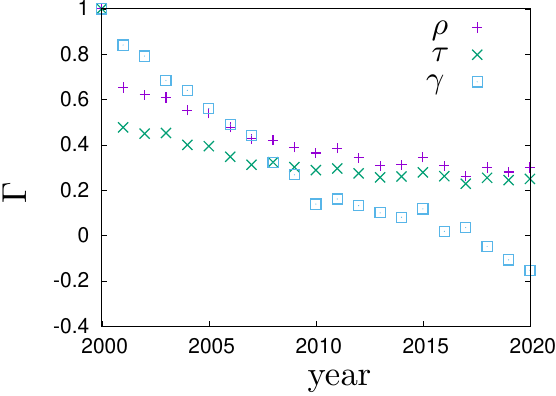}
\end{subfigure}
%% ------------------------------------------------------------------------------------
\begin{subfigure}[b]{0.49\columnwidth}
\caption{\label{fig:real_data_www_world}the world, data from \cite{Forbes}}
\vspace{2mm}
\includegraphics[width=0.99\columnwidth]{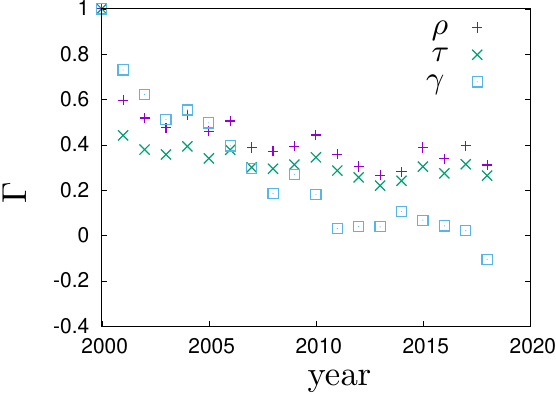}
\end{subfigure}
%% ------------------------------------------------------------------------------------
\caption{\label{fig:real_data}Time evolution of various rank correlations $\Gamma$ for top-100 richest people in (a) Poland, (b) Germany, (c) the USA and (d) the world.}
\end{figure}

In the remainder of this section we will investigate the rheology of wealth in real world systems. People's wealth data is very sensitive, so it is almost impossible to
collect it. Therefore we restrict ourselves to data on the richest people that is publicly available. 
To be specific, we focus on the top-100 lists
of richest people in Poland, Germany, the USA and the
world \cite{Wprost,Spiegel,Forbes}. The top-100 richest lists are a small part of the whole picture
but it is the part that usually gets the most attention.
Data that we analyse covers the period 2000--2020 for Poland \cite{Wprost}, Germany \cite{Spiegel} and the USA \cite{Forbes}, and the period 2000--2018 for the world \cite{Forbes}. 
We are going to determine the rank correlation coefficients $\tau$ \eqref{tau},
$\rho$ \eqref{rho}, and $\gamma$ \eqref{gamma} between the top-100 richest lists 
in years $2000$ and $2000+t$. The sets of people present in the top-100 richest lists 
vary from year to year: there are people who are in the top-100 richest list in some 
years but not in others. To compare the lists we must first standardize them so that they include the same set of people every year. To do so, we determine a full set of people who have 
been present on the top-100 list at least once in the studied period. The full set is then used to complete the annual top-100 lists in the following way:
if a person from the full set is not in the top-100 richest list for a given year, 
s/he is added to this list with a unique random rank in the range between $101$ 
and the number of people in the full set. In the analysed period, the full sets contain $360$ people in Poland, $342$ in Germany, $319$ in the USA and $323$ in the world. 
For each system, all standardized lists are of the same size and contain the same people every year. The standardized lists are used as input to calculations of $\tau$ and $\rho$. The results are shown in Figure \ref{fig:real_data}.
In the figure you can also see the results for Goodman--Kruskal's $\gamma$ (\ref{gamma}) computations. To compute $\gamma$ we used a slightly different algorithm to complement the annual lists. The algorithm assigns the same \emph{ex aequo} rank to all people added as a complement,  to the list.
By design, the coefficient $\gamma$ omits \emph{ex aequo} ranks, thus minimizing the statistical bias coming from the artificial completion of lists. 
Analysing the plots in Figure \ref{fig:real_data} we see for example, that the values of $\gamma$ drop between $k=0$ and $k=1$  much less than the corresponding values of $\tau$ and $\rho$. The big  drop of $\tau$ and $\rho$ is related to
the large number of additional rank pairs,
created by the complement algorithm. The number of pairs 
is less in the complement algorithm for $\gamma$.
When the sets of elements vary from year to year, the Goodman--Kruskal's coefficient $\gamma$ better captures rank correlations of the original ranking lists. Therefore, we find it more reliable to use $\gamma$ to compare rank correlations for lists of different lengths. 

%% ====================================================================================
%% ====================================================================================

\begin{figure}[ht]
\centering
%% ------------------------------------------------------------------------------------
\begin{subfigure}[b]{0.49\columnwidth}
\centering
\caption{\label{fig:real_data_overlap}}
\includegraphics[width=0.99\textwidth]{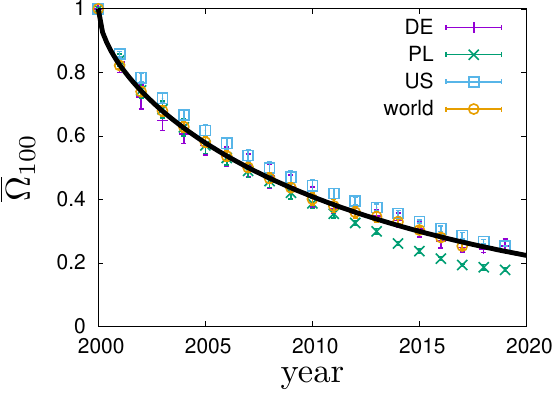}
\end{subfigure}
%% ------------------------------------------------------------------------------------
\begin{subfigure}[b]{0.49\columnwidth}
\centering
\caption{\label{fig:real_data_gamma}}
\includegraphics[width=0.99\textwidth]{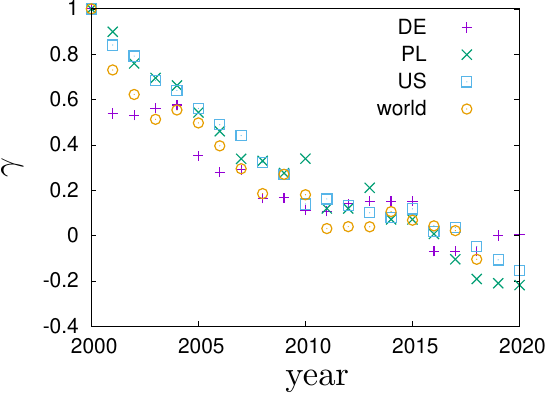}
\end{subfigure}
%% ------------------------------------------------------------------------------------
\caption{\label{fig:real_data_overlap_gamma}Rank coefficients for the richest people in Germany, Poland, the US and the world based on real-world data published in Refs. \cite{Wprost,Forbes,Spiegel}.
(a) Overlap ratios $\overline\Omega_{100}(t)$ for the four systems.
The best fit of the formula \eqref{eq:omega}, with $A=0.1536(42)$ and $B=0.0425(14)$, to the world data is shown with a solid line. (b) Goodman--Kruskal's $\gamma$ coefficients for the four system.}
\end{figure}

In Figure \ref{fig:real_data_overlap}, values of the overlap ratio of the annual top-100 richest lists in Poland, Germany, the US and the world are compared. The overlap ratios exhibit a fairly universal behaviour that is 
well described by the phenomenological formula
\eqref{eq:omega}. The best fit to the top-100
richest people in the world is shown with a solid
line in Figure \ref{fig:real_data_overlap}. It fits
the data very well. We can see that in the four systems
the overlap ratio drops to $50\%$ after
$7\pm 1$ years. In Figure \ref{fig:real_data_gamma}, values of the 
Goodman--Kruskal's rank correlation coefficient $\gamma$ for the four systems are compared. 
We can see that the rank correlations fall off from one to
zero roughly within $20$ years. The rate of decay for $\gamma$
is quite similar in all the studied systems. For Germany the dependence of $\gamma$ on time exhibits an interesting pattern. The values of $\gamma$ stay roughly constant for a couple of years, then significantly drop and again stay roughly constant for a couple of years. This motif repeats a couple of times, forming a stairway shape. 

%% ====================================================================================
%% Final set of parameters            Asymptotic Standard Error
%% =======================            ==========================
%% aom2            = 0.153556         +/- 0.004195     (2.732%)
%% bom2            = 0.0424926        +/- 0.001395     (3.283%)
%% ====================================================================================

%% ####################################################################################
\section{Conclusions}
%% ####################################################################################
Steady-state macroeconomic systems are characterized by no statistical changes in the distribution of wealth and macroeconomic parameters.
This does not mean that there are no changes in the system. On the contrary, in the microscale the system undergoes continuous dynamical changes. As far as the wealth distribution is concerned, 
the internal dynamics can be observed as a continuous process which manifests as
reshuffling of wealth ranks of people on the wealth ranking lists: some people get richer, 
some get poorer. In this paper we have studied the dynamical properties of macroeconomic systems related to the flow of wealth, and wealth rheology. We used the Bouchaud--M\'ezard model and the agent-based modelling approach to simulate the evolution of wealth in a closed macroeconomy. An interesting property of the Bouchaud--M\'ezard model
is that it generates steady states with a Pareto tail in the wealth distribution,
and that different combinations of economic activity parameters may lead to the same limiting wealth distribution. This allowed us to investigate the dependence 
of wealth rank correlations on the wealth activity parameters for various scenarios
but the same wealth distribution. We have seen that the wealth rank
correlations depend mainly on the growth rate volatility. The smaller the volatility, 
the larger the correlations. The rank correlations are closely
related to auto-correlations in the system. We have also studied rank correlations on 
the top-100 richest lists in Poland, Germany, the US and the world using the Goodman--Kruskal's $\gamma$ and the lists overlap ratio $\overline\Omega_{100}$. We have observed that the rank correlations in the studied
systems exhibit a certain degree of universality, for example, the overlap ratio gets reduced by half within $7\pm 1$ and the Goodman--Kruskal's $\gamma$ decreases to zero within two decades. We have focused on the top-100 richest lists because they are easily available. It would be interesting to perform a similar analysis for top-$n$ lists for larger $n$ $(n>100)$ and for longer periods, $t>20$ years, in the future to get a full picture on wealth rank dynamics. 

There are also some theoretical questions---regarding the scaling and universality of results---that are extremely interesting. We have seen that the correlation coefficients
$\rho$ and $\tau$ depend on a universal variable 
$k\alpha\sigma^2$ while the overlap ratio $\Omega_n$ on
$k(\alpha-1)\sigma^2$. The question is if this scaling
is the general feature of the Kesten processes?
To solve this problem we have considered an alternative implementation of Kesten dynamics, where the wealth of agents is generated as exponents $W_{a,k} = \exp(x_{a,k})$ of independent random walks $x_{a,k}$, $a=1,\ldots, N$ on the positive semi-axis with the drift towards the reflective wall located at $x=0$. Such a stochastic process is probably the simplest implementation of the Kesten process. It leads to
a stationary state with a power law in the probability density function of wealth distribution
$p(W) = \alpha W^{-1-\alpha}$ for $W\ge 1$, with the exponent $\alpha = -\mu/\sigma^2$, where $\mu,\sigma$ are the drift and the volatility of the underlying random walk. The rank statistics of $W$'s and $x$'s are the same because the map $x\rightarrow \exp(x)$ is monotonic. The Kesten dynamics is thus mapped onto the dynamics of a gas of particles performing independent random walks on the positive semiaxis. The gas of particles is much easier to analyse.
For instance, one can give a simple argument that the overlap ratio for this system depends on the combination 
$k\sigma^2 \alpha^2$, so the scaling is different than
for the model discussed in this paper. Thus,
the scaling is not universal. It would be interesting
to find an analytical way to calculate the rank correlation
measures in the models discussed in this paper. For the
gas of random walkers on the semi-positive axis one 
can maybe use the spectral methods \cite{KmG}. 
%% ####################################################################################

\authorcontributions{All authors have contributed equally to this work.
%% Conceptualization, X.X. and Y.Y.; 
%% methodology, X.X.; 
%% software, X.X.; 
%% validation, X.X., Y.Y. and Z.Z.; 
%% formal analysis, X.X.; 
%% investigation, X.X.; 
%% resources, X.X.; 
%5 data curation, X.X.; 
%% writing--original draft preparation, X.X.; 
%% writing--review and editing, X.X.; 
%% visualization, X.X.; 
%% supervision, X.X.; 
%% project administration, X.X.; 
%% funding acquisition, Y.Y.
They have read and agreed to the published version of the manuscript.
%%'', please turn to the  \href{http://img.mdpi.org/data/contributor-role-instruction.pdf}{CRediT taxonomy} for the term explanation. Authorship must be limited to those who have contributed substantially to the work reported.
}

%%%%%%%%%%%%%%%%%%%%%%%%%%%%%%%%%%%%%%%%%%
\funding{MS has been partially supported by the Ministry of Education and Science within the ``Regional Initiative of Excellence'' Program for 2019--2022.}

%%%%%%%%%%%%%%%%%%%%%%%%%%%%%%%%%%%%%%%%%%
\acknowledgments{We thank the anonymous reviewer who drew our attention to the scaling of dynamics in the model.}

%%%%%%%%%%%%%%%%%%%%%%%%%%%%%%%%%%%%%%%%%%
\conflictsofinterest{The authors declare no conflict of interest.}

%%%%%%%%%%%%%%%%%%%%%%%%%%%%%%%%%%%%%%%%%%
%% optional
%%\abbreviations{The following abbreviations are used in this manuscript:\\
%%\noindent 
%%\begin{tabular}{@{}ll}
%%MDPI & Multidisciplinary Digital Publishing Institute\\
%%DOAJ & Directory of open access journals\\
%%TLA & Three letter acronym\\
%%LD & linear dichroism
%%\end{tabular}}

%%%%%%%%%%%%%%%%%%%%%%%%%%%%%%%%%%%%%%%%%%
%% optional
\appendixtitles{yes} 
\appendixstart
\appendix

%% ####################################################################################
\section{\label{AppA}Rank correlations}
%% ####################################################################################

Consider a set of $N$ elements, each being characterised by two real quantities $X_i$, $Y_i$,  
$i=1,\ldots,N$. For example one can think of the weight and height 
of pupils in a class. The elements (pupils) can be ranked with respect 
to the property $X$ (weight) and $Y$ (height).
For sake of simplicity assume that all values $X_i$, $i=1,\ldots, N$ are distinct, so that the elements can be unambiguously ranked with respect to $X$, and assume that the same holds for $Y$. 
Denote the corresponding ranks by $r_i$'s and $s_i$'s. To quantify correlations between the ranks one can introduce a rank score 
function $f$ to define a rank correlation coefficient associated with
this score function
\begin{equation}
\Gamma_f = \frac{\sum_{i,j=1}^N  f(r_i-r_j) f(s_i-s_j)}
{\sqrt{\sum_{i,j=1}^N f^2(r_i-r_j)}\sqrt{\sum_{i,j=1}^N f^2(s_i-s_j)}}.
\end{equation}
The score function is a monotonic odd function $f(-x)=-f(x)$. 
The standard choices of the rank score function are 
the sign function $f(x) = \sign(x)$ or
the identity function $f(x)=\id(x)=x$. In the former case the 
corresponding rank correlation
coefficient is usually called Kendall's $\tau$.
It is calculated as follows:
\begin{equation}
\tau  \equiv  \Gamma_{\sign} = \frac{2 \sum_{i,j=1}^N 
 \sign(r_i-r_j){\sign} (s_i-s_j)}{N(N-1)} = \frac{2(N_c-N_d)}{N(N-1)},
 \label{tau}
\end{equation} 
where $N_c$ and $N_d$ are the numbers of concordant and discordant pairs.
A pair $(i,j)$ is concordant if the elements $i$ and $j$ are ranked in the same order for $X$ and $Y$. Otherwise, the pair is discordant.
For the identity function the rank correlation coefficient 
is usually called Spearman's $\rho$
\begin{equation}
\rho \equiv \Gamma_{\id}  = \frac{\sum_{i,j=1}^N (r_i-r_j) (s_i-s_j)}{\sum_{i,j=1}^N (r_i-r_j)^2}.
\end{equation} 
Using elementary algebraic operations one can find that
\begin{equation}
\rho  = 1 - \frac{6 \sum_{i=1}^N (r_i-s_i)^2}{N(N^2-1)}.
\label{rho}
\end{equation} 
This formula is particularly useful in practice since instead of a double sum it involves
only a single sum. The computational complexity to determine $\rho$ increases linearly with $N$ in contrast to the complexity to compute $\tau$ \eqref{tau} which increases quadratically with $N$.

We conclude this Appendix by mentioning that there are rank correlation measures that apply also for the case when there are \emph{ex aequo} ranks 
in the sample. In the example given above, you can think of two or more 
pupils having the same weight or height within a given accuracy. 
In this case, one can replace Kendall's $\tau$ by Goodman--Kruskal's $\gamma$ \cite{Goodman}
\begin{equation}
\gamma=\frac{N_c-N_d}{N_c+N_d}.
\label{gamma}
\end{equation}
Clearly, if there are no \emph{ex aequo} ranks in the sample then 
$\gamma$ is equivalent to $\tau$ \eqref{tau}.

 %%%%%%%%%%%%%%%%%%%%%%%%%%%%%%%%%%%%%%%%%%

\end{paracol}                      

%% ####################################################################################
\reftitle{References}
%\externalbibliography{yes}
%\bibliography{bibliography}

\end{document}